\documentclass[aps,pra,showpacs,superscriptaddress]{revtex4}
\usepackage{graphicx}
\usepackage{dcolumn}
\usepackage{bm}
\usepackage{color}
\usepackage{amsmath}

\begin{document}

\title{All-optical 3D atomic loops generated with Bessel light fields}
\author{Karen Volke-Sep\'ulveda}
\email{karen@fisica.unam.mx}
\author{Roc\'{\i}o J\'auregui}
 \email{rocio@fisica.unam.mx}
 \affiliation{Departamento de  F\'{\i}sica
Te\'orica, Instituto de F\'{\i}sica, Universidad Nacional Aut\'onoma
de M\'exico, A.P. 20-364, M\'exico 01000 D.F. M\'exico}

\date{\today}

 \begin{abstract}
 The propagation invariance of Bessel beams as well as their
 transversal structure are used to perform a comparative analysis
 of their effect on cold atoms for four different configurations
 and combinations thereof.
 We show that, even at temperatures for which the classical
 description of the atom center of mass motion is valid, the
 interchange of momentum, energy and orbital angular momentum
 between light and atoms yields efficient tools for all-optical
 trapping, transporting and, in general,
 manipulating the state of motion of cold atoms.
 \end{abstract}
 \maketitle

\section{Introduction}

Over the last decade, experiments on the interaction between light
and cold atoms have seen tremendous advances. Laser cooling of
neutral atoms is nowadays a well-established procedure and  solid
steps for novel experiments in research areas such as atom optics
and quantum information processing with atomic systems have been
taken. During the last fifteen years, the development of far off resonance
traps (FORT's) \cite{fort} has allowed the organization of cold
matter in optical lattices \cite{opt latt} and, with this, the study
of single particle Bloch physics. The creation of trapped degenerate
atomic gases, on the other hand, is one of the most
exciting scientific achievements of modern times \cite{BEC}, as it has
opened, for instance, the possibility of realizing interference of
matter waves \cite{matter interf}.

In these areas, the use of light beams with especial intensity
and/or phase structure yielding peculiar dynamical properties, plays
a very important role. Bessel beams \cite{durninME}, for instance,
have been proposed as waveguides for atom transport due to their
propagation invariance \cite{ArltHD00,ad1}. The measurement of
the mechanical properties of Mathieu beams  could
be performed through the analysis of their effects on cold atoms \cite{blas}.
The transfer of orbital angular momentum (OAM) from
Laguerre-Gaussian (LG) laser modes \cite{allen92, luciana} or
high-order Bessel beams \cite{OAMBB, cojoc, twisted light} to cold
matter has been the subject of theoretical studies in both, paraxial and
non-paraxial regimes. For the case of a LG beam interacting with a
diatomic molecule, it was found that OAM has in general a weak
effect on the internal state, since it becomes relevant at the electric
quadrupole interaction level, while the major mechanism of exchange
occurs in the electric dipole approximation and involves only the
center of mass motion \cite{luciana}. For non-paraxial Bessel beams
interacting with a single atom, in contrast, the probability that
the internal state of an atom acquires orbital angular momentum from
light is maximum when the atom is located at the beam axis
\cite{twisted light}. In fact the helicity factor $k_zc/ \omega$,
which is related to the projection of angular momentum along  the
main direction of propagation, could be used to directly enhance or
suppress atomic transitions \cite{mechanical}.

The first experimental demonstration of OAM transfer to cold atoms
was reported almost a decade ago by Tabosa and coworkers
\cite{TabosaP99} and, very recently, OAM was transferred to a Bose
Einstein condensate \cite{OAM BEC}. OAM  transfer is also an
important aspect in the study of circular optical lattices and
helical waveguides, which are interesting alternatives for
interference experiments with matter waves and quantum transport.
For example, Haroutyunyan and Nienhuis \cite{nienhuis} have recently
explored the use of stationary waves in the angular direction,
generated by the superposition of two counter rotating LG beams
propagating in the same direction, as a more efficient alternative
for achieving the exchange of angular momentum between light and
cold atoms. In this case, the confinement in the radial direction is
achieved through an extra trapping potential with cylindrical
symmetry, but the dynamics along the $z$ axis is completely free. A
circular optical potential of this kind would split the wave
function of a single localized atom into clockwise and anticlockwise
components, which may interfere under certain confinement conditions
\cite{nienhuis}. Bhattacharya \cite{twist} on the other hand,
presented a simplified analysis of a curved helical lattice as an
atom guide, which could be generated by the superposition of two
identical LG beams propagating in opposite directions. This
motivated, in turn, a study on the possibility of having bound
states in such curved helical potentials \cite{exner}. Circular and
rotating optical lattices have been studied as well, in the context
of condensed matter and many particle systems, such as Fermi gases
and Bose Einstein condensates \cite{bec current}-\cite{rot OL}.
Moreover, ring shaped optical lattices represent appropriated
potentials for studying quasi one dimensional physical systems with
closed boundary conditions \cite{amico, olson07}. Most of these
systems consider, besides optical fields, external magnetic fields
to achieve confinement in one or more spatial dimensions.

In this work we perform a theoretical comparative analysis of the
effect of four different light fields of circular structure
on a dilute gas of cold atoms whenever the
effect of collisions among them  can be neglected. In all cases, we
use Bessel beams (BB) and superpositions thereof, taking advantage
of their propagation invariance property. First, we analyze the case
of a single high order BB. Secondly, we study the case
of an optical field with $2m$ intertwined helicoidal lines of light,
similar to the curved
helical waveguides studied by Bhattacharya \cite{twist}, but in this
case, it results from superimposing two identical BBs propagating in
opposite directions. In the third place, we look at a
three dimensional circular lattice, corresponding to the
simultaneous generation of standing waves in the radial, angular and
axial directions.  As a fourth option, we analyze a circular
optical lattice constituted by a collection of individual toroidal
traps along the $z$-axis, which can be achieved by interfering two
counter propagating BBs of opposite helicity. Our interest in these
particular configurations arises from the fact that they can be
combined and used successively for creating  \textquoteleft atom
loops'  in predesigned ways, as it will be demonstrated
in the last section of the paper. Our approach  follows the
semiclassical description made by Gordon and Ashkin \cite{gordon-ash},
which can be applied when the atom velocity is sufficiently low but
not beyond the quantum limits. The quantum mechanics treatment will
be presented elsewhere.

\section{Rotating light beams and circular lattices}

As a starting point for the present discussion, we will briefly
describe the main properties of a BB within a vectorial treatment,
in order to account for polarization properties. Under ideal
conditions, the electromagnetic field of a Bessel mode has
cylindrical symmetry which guarantees its propagation invariance
along the $z$-axis:
\begin{eqnarray}
E_{\rho }^{\kappa }&=&\frac{\partial ^{2}}{\partial z\partial \rho
}\Pi _{\kappa }^{(TM)}-\frac{1}{\rho }\frac{\partial ^{2}}{c\partial
t\partial
\phi }\Pi _{\kappa }^{(TE)},\\
E_{\phi }^{\kappa }&=&\frac{1}{\rho }\frac{%
\partial ^{2}}{\partial z\partial \phi }\Pi ^{(TM)}_\kappa+\frac{\partial ^{2}}{%
c\partial t\partial \rho }\Pi _{\kappa }^{(TE)},\\
 E_{z}^{\kappa }&=&-\frac{%
\partial ^{2}}{c^{2}\partial t^{2}}\Pi _{\kappa }^{(TM)}+\frac{\partial ^{2}%
}{\partial z^{2}}\Pi _{\kappa }^{(TM)},
\end{eqnarray}
where $\Pi _{\kappa }^{j}=\mathcal{E}_{\kappa }^{j}\psi _{m}(k_{\bot }\rho
,\phi )e^{i(k_{z}z-\omega t)}$, with $\psi _{m}(k_{\bot }\rho ,\phi
)=J_{m}(k_{\bot }\rho )e^{im\phi }$. $J_{m}$ is the Bessel function
of order $m$ and $\mathcal{E}^{(TE)}$ ($\mathcal{E}^{(TM)}$) is proportional to
the amplitude of the transverse electric (magnetic) mode. $\kappa $ denotes
collectively the parameters that define the mode, that is, the propagation
wavenumber along the $z$ axis, $k_{z}$, the transverse propagation
wavenumber $k_{\bot }$ and the azimuthal index $m$.

By superpositions of TE and TM Bessel modes, different polarizations states
can be obtained. In the literature \cite{OAMBB,barnett94, kvlk06} the modes
\begin{eqnarray}
\vec E_{m}^{(\mathcal{L})}(\vec r,t;k_{\bot },k_{z}) &=&E_{0}^{(%
\mathcal{L})}\Big[(\hat e_x + i \hat e_y)\psi _{m}-\frac{i}{2}\Big(%
\frac{k_{\bot }}{k_{z}}\Big)\psi _{m+1}\hat e_z \Big],  \notag \\
\vec E_{m}^{(\mathcal{R})}(\vec r,t;k_{\bot },k_{z}) &=&E_{0}^{(%
\mathcal{R})}\Big[(\hat e_x-i\hat e_{y})\psi _{m}+\frac{i}{2}\Big(%
\frac{k_{\bot }}{k_{z}}\Big)\psi _{m-1}\hat e_{z}\Big],
\label{eq:circpol}
\end{eqnarray}
are considered to be the analogues of left ($\mathcal{L}$) and
right-handed ($\mathcal{R}$) circularly polarized plane wave modes.
Their superpositions
 $\mathbf{E}_{m}^{(\mathcal{R})}\pm \mathbf{E}_{m}^{(\mathcal{L})}$ define
linearly polarized modes. In terms of TE and TM modes,
\begin{eqnarray}
\vec E_{m}^{(\mathcal{L})} &=&E_{0}^{(\mathcal{L})\prime }\Big(\vec E%
_{m+1}^{(TM)}-i\frac{ck_{z}}{\omega }\vec E_{m+1}^{(TE)}\Big)  \notag \\
\vec E_{m}^{(\mathcal{R})} &=&E_{0}^{(\mathcal{R})\prime }\Big(\vec E%
_{m-1}^{(TM)}+i\frac{ck_{z}}{\omega }\vec E_{m-1}^{(TE)}\Big).
\label{eq:rl}
\end{eqnarray}

The mechanical properties of the photons associated with Bessel modes
are directly related to the numbers $\omega$, $k_{z}$, $m$ that
characterize them, along with the polarization. In fact, $\hbar
\omega$, $\hbar k_{z}$, $m\hbar $ correspond to the energy, linear
momentum and orbital angular momentum along the $z$ direction
respectively \cite{pol}.   A linearly polarized mode has the
structure
\begin{equation}
\vec E_{m}(\vec r,t;k_{\bot },k_{z})=E_{0}\Big[(\psi _{m}\hat e_{x}-\frac{i}{2}\Big(\frac{k_{\bot }}{%
k_{z}}\Big)(\psi _{m+1}-\psi _{m-1})\hat e_{z}\Big]. \label{Pstate}
\end{equation}

In what follows, we will describe in some detail each one of the
four optical fields of interest, providing an explicit analytical
expression and a brief discussion about their experimental
generation. For this purpose, it will be useful to establish first a
distinction between rotating and stationary BBs.

While a linearly polarized rotating BB corresponds to that given by
Eq.(\ref {Pstate}), with $\psi _{m}(\rho ,\varphi )=J_{m}(k_{\bot
}\rho )e^{im\varphi }$, a stationary BB is formed by the
superposition of two rotating BBs of the same topological charge
$\left| m\right| $ traveling along the same axis and direction, but
rotating in opposite sense ($\pm m$), giving rise to
\begin{eqnarray}
&&\vec E_{m}(\vec r,t) =E_{0}e^{i\left(
k_{z}z-\omega t-\varphi_0\right)}\Big[ J_{m}(k_\bot\rho)\cos (m\varphi +\varphi_0
)\hat e_{x}\nonumber\\
&&-\frac{i}{2} \frac{%
k_{\bot }}{k_{z}} \Big( J_{m+1}(k_\bot\rho)\cos [ (m+1)\varphi +\varphi_0 ]
-J_{m-1}\cos [ (m-1)(k_\bot\rho)\varphi +\varphi_0] \Big) \hat e_{z}\Big].
\label{stat}
\end{eqnarray}
Here and in the following, $\varphi_0 =0$ ($\varphi_0 =\pi
/2$) stands for even (odd) values of $m$. Figure~1(a) illustrates
the intensity and phase distributions of an ideal rotating BB, while
Figure~1(b) shows the same for an ideal stationary BB, in both cases
$m=2$. It is seen that the ideal fields exhibit propagation
invariance of their intensity along the $z$ axis.
\begin{figure}
\includegraphics[width=3in]{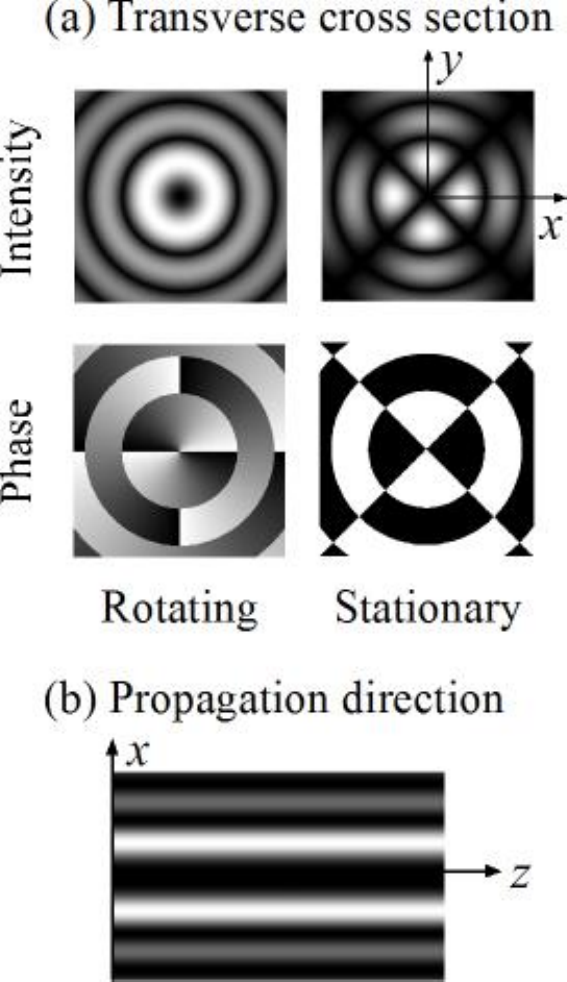}
\caption{Comparison between (a) a rotating and (b) a stationary Bessel beam
 of order $m=2$.
The phase values for both beams are indicated in the color bar on
the right hand side, in units of $\pi$ radians. Along the
propagation axis, both beams have a similar behavior due to their
ideal propagation invariance.} \label{fig1}
\end{figure}

The optical fields of interest will be constructed as superpositions of
linearly polarized Bessel modes in the sense discussed above, either
rotating or stationary, with the transverse component of their electric
fields oriented along the $x$ axis.\\

\vspace{0.5cm}
{\bf Case 1}. Single rotating Bessel beam.

A single BB in interaction with cold atoms has been studied before
\cite{ArltHD00, ad1,twisted light}. Here we include this simple case
for comparative purposes with the other configurations and also to
emphasize some of its applications for controlling atomic motion.
The expression for a linearly polarized rotating BB is, according to
Eq.(\ref{Pstate}),

\begin{equation}
\vec E_{m}^{(1)}(\vec r,t;k_{\bot },k_{z})=E_{0}e^{ i\left( k_z
z+m\varphi -\omega t\right)}\Big[ J_{m}(k_{\bot} \rho)
 \hat e_{x}-\frac{i}{2}\frac{k_{\bot }}{%
k_{z}}\Big( J_{m+1}(k_{\bot} \rho)e^{i\varphi}- J_{m-1}(k_{\bot}
\rho)e^{-i\varphi} \Big)\hat e_{z}\Big]. \label{1st}
\end{equation}
Experimentally, reasonable approximations to BBs of different orders
have been efficiently generated by illuminating an axicon or conical
lens with a single-ringed Laguerre-Gaussian mode of order $m$
\cite{ArltD00}. Another approach is to obtain the desired BB
directly from properly designed computer generated holograms (CGH)
\cite{vasara}, which can be displayed in spatial light modulators (SLM)
\cite{bb-slm}. The original setup proposed by Durnin and coworkers
\cite{durninME}, consisting of a dark screen with a thin annulus
transmitance function placed at the back focal plane of a positive
lens, turns out to be inefficient for optical trapping experiments,
although it is the best approximation to the theoretical expression.
In all cases, of course, BBs can be generated only within a finite
region and, under current experimental conditions, the paraxial
approximation is generally fulfilled. It is worth to mention,
however, that BBs with relatively large transverse dimensions
($k_{\bot}/k_z << 1$) can be reduced with additional lenses in order
to make them more suitable for atom trapping experiments.\\

\vspace{0.5cm}
{\bf Case 2}. Twisted helical lattice.

\begin{figure}
\includegraphics[width=3in]{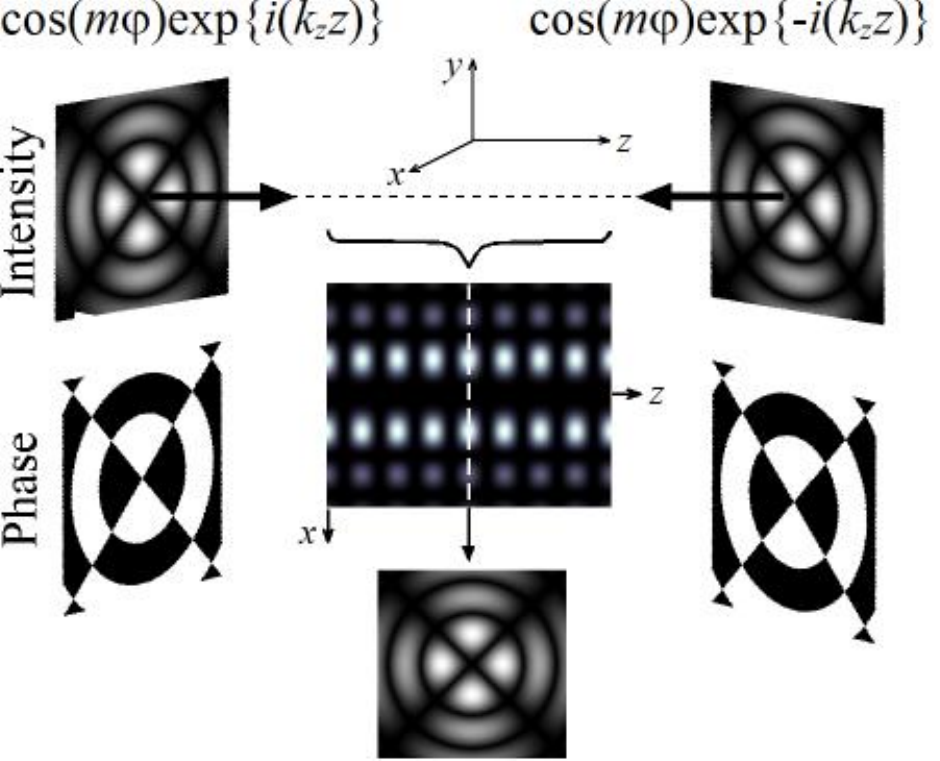}
\caption{(Color online) Schematic of the superposition of two
rotating BBs propagating in opposite directions; the rotation sense
of the beams is opposite as well with respect to the same fixed
reference frame. Transverse cross sections of the resulting field at
different $z$ planes indicate an intensity distribution that is
twisted around the $z$ axis, but stationary in time. The whole
structure resembles a rope with $2m$ main inner strands of light
twisted together, and outer groups of strands with reduced
intensity.}\label{fig2}
\end{figure}

This field can be generated by the interference of two rotating BBs
with the same helicity but propagating in opposite directions. This
means that the two beams have the same projection of their
respective angular momenta along their own propagation direction
but, with respect to the same and fixed reference frame, they are
rotating in opposite directions, as illustrated in Fig.~\ref{fig2}.
The resulting field is described by

\begin{eqnarray}
\vec E_{m}^{(2)}(\vec r,t) &=&E_{0}e^{-i(\omega
t+\varphi_0)}\Big[ J_{m}(k_{\bot }\rho )\cos (m\varphi +k_{z}z+\varphi_0 )\hat
e_{x}\nonumber\\ &-&\frac{i}{2}\left( \frac{k_{\bot
}}{k_{z}}\right)
\Big( J_{m+1}(k_\bot\rho )\cos \left[ (m+1)\varphi +k_{z}z+\varphi_0 %
\right] - J_{m-1}(k_\bot\rho )\cos \left[ (m-1)\varphi
+k_{z}z+\varphi_0 \right] \Big) \hat e_{z}\Big] . \label{3rd}
\end{eqnarray}
The experimental generation of this optical field can be performed
by introducing a rotating BB into an amplitude division
interferometer; each portion of the split beam should suffer the
same number of reflections, so that the helicity is preserved for
both of them before being superimposed again along the same axis
while propagating in opposite directions.\\

\vspace{0.5cm}
{\bf Case 3}. 3D stationary circular lattice.

In this case, we consider the interference of two stationary Bessel modes
of the type described by Eq.~(\ref{stat}), but propagating in opposite
directions along the same $z$ axis. The resulting optical field will exhibit
standing waves in all the three spatial dimensions within a circular cylindrical geometry,

\begin{eqnarray}
\vec E_{m}^{(3)}(\vec r,t) &=&E_{0}
e^{-i(\omega t+\varphi_0)}\cos (k_{z}z)\Big[J_{m}(k_{\bot }\rho )\cos (m\varphi
+\varphi_0 ) \hat e_{x}\nonumber\\
&-&\frac{i}{2}\frac{k_{\bot }}{k_{z}}  \Big(
J_{m+1}(k_\bot\rho )\cos \left[ (m+1)\varphi +\varphi_0
\right]-J_{m-1}(k_\bot\rho )\cos \left[ (m-1)\varphi +\varphi_0 \right]\Big) \hat e%
_{z} \Big] .  \label{2nd}
\end{eqnarray}

Intensity nodal  surfaces correspond, along the
 radial direction, to concentric dark cylinders
 whose radii $\rho=\rho_{mn}$, are defined by $k_\bot\rho_{mn}=x_{nm}$,
 with $x_{mn}$ the $n$-th root of the Bessel function of order $m$.
 The cylinders are intersected by $2\left|m\right|$ semi-infinite
 nodal planes along the azimuthal coordinate,
 defined by $\left[\left|m\right|\varphi_n+\varphi_0\right]=(2n-1)\pi/2$,
 where $n=1,2,...,2\left|m\right|$. Finally, there are also nodal planes
 along the $z$ axis corresponding to $z_n=(2n-1)\lambda_z/4$, with $n$
 an integer and $\lambda_z=2\pi/k_{z}$.
\begin{figure}
\includegraphics[width=3in]{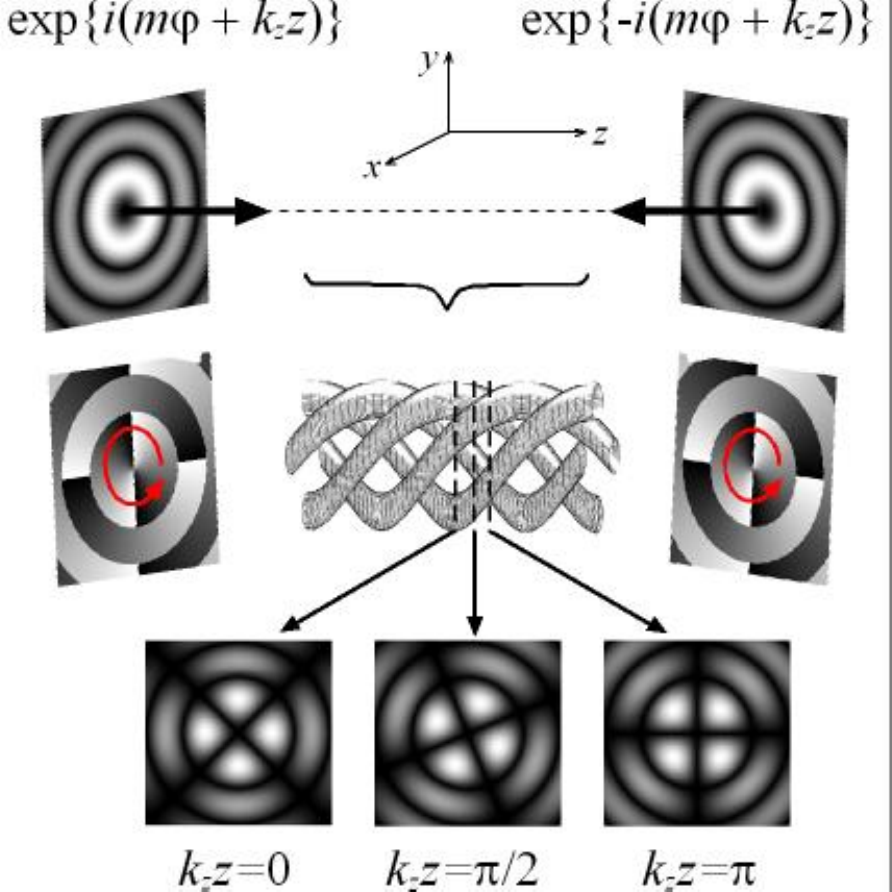}
\caption{Schematic of the superposition of two stationary BBs
propagating in opposite directions. Standing waves are generated in
the three spatial directions: radial, angular and
axial.}\label{fig3}
\end{figure}
The experimental generation of a lattice like this may involve two
steps. First, it is necessary to obtain the stationary BB, which can
be done either directly, by means of CGH, or by interfering two
counter rotating BBs propagating along the same axis and direction,
for instance. Once obtained the stationary BB, an amplitude division
interferometer would be appropriated for superimposing two equally
weighted portions of it, aligned along the same axis, but
propagating in opposite directions. This is schematically
illustrated in Fig.~\ref{fig3}.\\

\vspace{0.5cm}
{\bf Case 4}. Toroidal train lattice.

A set of toroidal traps along an axis can be generated by the
interference of two rotating BBs with opposite helicities and
propagating in opposite directions. This means that the two beams
have opposite projections of their respective angular momenta along
their own propagation direction but, with respect to the same
reference frame, they are rotating in the same direction, as
illustrated in Fig.~4. The resulting field is given by

\begin{eqnarray}
\vec E_{m}^{(4)}(\vec r,t) &=&E_{0}e^{ i\left( m\varphi
-\omega t\right)} \cos (k_{z}z)\Big[ \hat e_{x}J_{m}(k_{\bot
}\rho )\nonumber \\
&-&\frac{i}{2}\left( \frac{k_{\bot }}{k_{z}}\right)
\left[ J_{m+1}(k_\bot\rho )e^{i\varphi}-J_{m-1}(k_\bot\rho )e^{-i\varphi}\right] \hat
e_{z}\Big] . \label{4th}
\end{eqnarray}
A transverse cross section of this field at an antinodal
 plane along the $z$ axis
is exactly the same that the transverse cross section of a
propagating BB, but null intensity occurs at the $z$ nodal planes
 which, as in Case 3, correspond to the planes $z=(2n-1)\lambda_z/4$ ($n$ integer).
The experimental generation of this optical field can be performed
by introducing a rotating BB into an amplitude division
interferometer; one portion of the split beam should suffer an extra
reflection, so that its helicity is inverted with respect  to the
other portion of the beam before joining them together along the
same axis while propagating in opposite directions.

As we shall see in the next section, Bessel optical modes, either stationary
or propagating, exhibit interesting features in interaction with cold atoms,
due to their propagation invariance property and their multiringed radial
structure.
\begin{figure}
\includegraphics[width=3in]{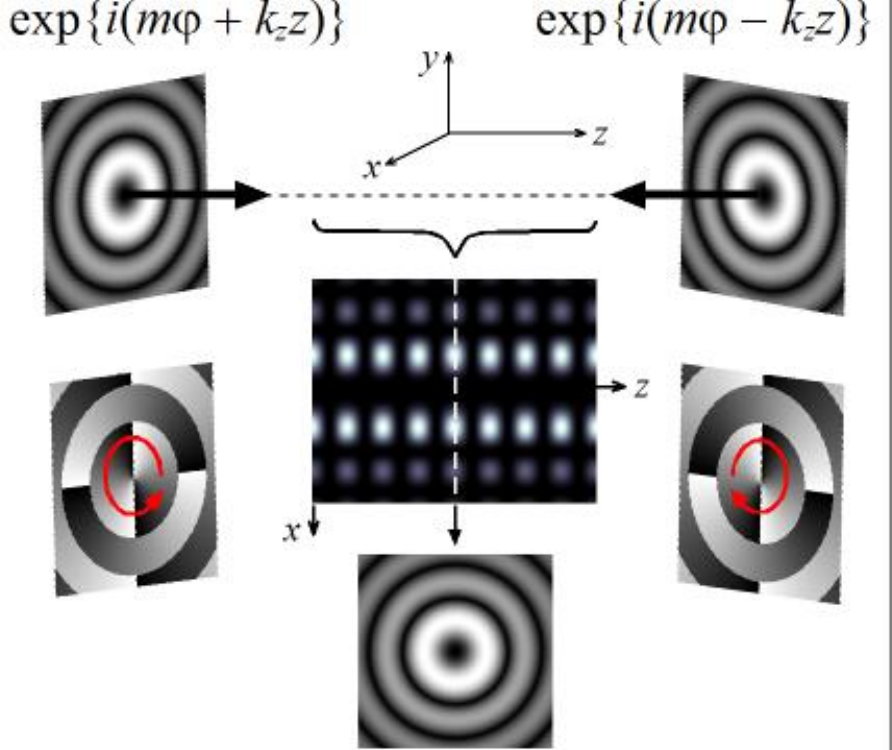}
\caption{(Color online) Schematic of the superposition of two
rotating BBs propagating in opposite directions; the rotation sense
of the beams in this case is the same with respect to the same fixed
reference frame. The 3D intensity distribution would resemble a
straight backbone of light.}\label{fig_4}
\end{figure}

\section{Semiclassical description of a single atom motion within the light field}

We take the standard semiclassical description as in the pioneer
works by Letokhov and coworkers \cite{leto} and Gordon and Ashkin
\cite{gordon-ash}. In this approximation, a monochromatic
electromagnetic wave describable by a coherent state couples to the
dipole moment of an atom. This dipole moment $\vec \mu_{12}$ is
related to the electromagnetic transitions between the atom levels
that, for simplicity, will be taken to have just two accessible
options. The coupling $g=i\vec \mu_{12}\cdot \vec E/\hbar$, depends
explicitly on the orientation of the electric field $\vec E$ of the
wave. For the systems described in this work, $\vec E$ arises from
linearly polarized beams and has a longitudinal component that is
much smaller than the transverse component since $k_z\gg k_\bot$. As
a consequence, from now on, the effect of longitudinal fields on the
atom will be neglected, i.e., $g\sim i\vec \mu_{12}\cdot \vec
E_\bot/\hbar=[i(\vec \mu_{12}\cdot {\bf e}_x) E_0/\hbar]\tilde
g(\rho,\varphi,z)$, where $\tilde g$ contains the information about
 the spatial structure of the light field.

In the following, we shall describe the atom motion under the
assumption that its kinetic energy is low enough to be sensitive to
the optical force but large enough to admit a classical description
in terms of Newton equations. The expression for the average
semiclassical velocity-dependent force \cite{gordon-ash}, valid for
both propagating and standing beams, is:
\begin{equation}
\langle \vec f \rangle = \hbar \tilde\Gamma p^\prime\Big[[(\vec
v\cdot\vec\alpha)(1-p)(1+p)^{-1} +\Gamma/2] \vec  \beta + [(\vec
v\cdot \vec\beta) -
\delta\omega]\vec\alpha\Big], \label{eq:force}
\end{equation}
with
\begin{equation}
\tilde\Gamma = \Gamma/[\Gamma(1+p^\prime) +2\vec
v\cdot\vec\alpha[1-p/p^\prime -p][p^\prime/(1+p)]],
 \label{eq:gtilde}
\end{equation}
$\Gamma =
4k^3\vert\vec\mu_{12}\vert^2/3\hbar$ the Einstein coefficient,
 $\delta\omega = \omega -\omega_0$ the detuning
between the wave frequency $\omega$ and the transition frequency
$\omega_0$, $p =2\vert
g\vert/^2((\Gamma/2)^2+\delta\omega^2)$ a parameter linked to the
difference $D$ between the populations of the two levels of the
atom,
 $D=1/(1+p)$, and finally $p^\prime = 2\vert g\vert^2/\vert \gamma^\prime\vert^2$, with
$\gamma^\prime =(\vec v\cdot\vec\alpha)(1-p)(1+p)^{-1} =\Gamma/2 +
i(-\delta\omega +(\vec v\cdot\vec\beta))$. The spatial structure of
a beam enters basically in the coupling factor $g=\vert g\vert e^{i
\phi}$, which appears in $p$ and $p^\prime$, and defines the vectors
$\vec \alpha$ and $\vec \beta$ through the equation $\nabla
g=(\vec\alpha
 + i\vec\beta)g$, leading to
 $\vec\alpha=\vec\nabla\ln \vert g\vert$ and $\vec\beta=\vec\nabla\phi$.

In experiments with cold atoms, it is well known that gravity
effects should, in general, be taken into
account to  describe accurately their motion. Here, we will consider
that the $z$ axis of the light field
configurations is oriented along the vertical direction. The atoms
are downloaded to the optical trap, with most of their kinetic
energy coming from the axial velocity which is assumed, unless
otherwise stated, to be negative.

In Table I, the spatial structure factor $\tilde
g(\rho,\varphi,z)$ associated with the different beam configurations
is given along with the force factors $\vec\alpha$ and $\vec\beta$.

\begin{table}
\begin{tabular}{|c|c|c|c|}
\hline Case& $\tilde g$ &$\vec \alpha$& $\vec \beta$\\
\hline 1& $J_m(k_\bot \rho)e^{i(k_zz-m\varphi)}$& $ k_\bot
\frac{J_m^\prime(k_\bot\rho)}{J_m(k_\bot\rho)}\hat e_\rho$ &
$\frac{m}{\rho}\hat e_\varphi+k_z\hat e_z$\\
\hline 2&$J_m(k_\bot \rho)\cos(k_zz+m\varphi)$& $ k_\bot
\frac{J_m^\prime(k_\bot\rho)}{J_m(k_\bot\rho)}\hat
e_\rho-(\frac{m}{\rho} \hat e_\varphi+k_z\hat e_z) \tan (m\varphi
+k_zz)$ &
$\vec 0$\\
\hline 3&$J_m(k_\bot \rho)\cos(k_zz)\cos (m\varphi)$& $ k_\bot
\frac{J_m^\prime(k_\bot\rho)}{J_m(k_\bot\rho)}\hat e_\rho-
\frac{m}{\rho}\tan (m\varphi)\hat e_\varphi-  \tan(k_zz)\hat e_z$ &
$\vec 0$\\
\hline 4&$J_m(k_\bot \rho)\cos(k_zz)e^{im\varphi}$& $ k_\bot
\frac{J_m^\prime(k_\bot\rho)}{J_m(k_\bot\rho)}\hat e_\rho- k_z \tan
(k_zz)\hat e_z$ &
$\frac{m}{\rho}\hat e_\varphi$\\
\hline
\end{tabular}\caption{ Coupling factor $\tilde g$, and conservative $\vec \alpha$
and dissipative $\vec \beta$ vectors defining the atom-BB
interaction, Eq.(\ref{eq:force}) for the different beam
configurations described in section 2.}
\end{table}
\bigskip

\section{Numerical results}

In this work we shall consider red detuned far off resonance light
beams. The bright regions of the light intensity distribution
correspond to minima of the effective potential energy $V_{eff}$
associated with the term $\vec \alpha=-\vec \nabla
V_{eff}=\vec\nabla\ln \vert g \vert$. The behavior of the atom in
the light field depends not only on its initial balance between
kinetic and effective potential energy, but also on its initial
momentum and position. In all the studied cases, we will
illustrate the behavior of an atomic cloud, which means that we will
show the paths of several atoms whose initial conditions vary within
a certain range of experimentally accessible values.

Unless otherwise stated, the parameters in the numerical simulations
consider $^{85}$Rb atoms. Following Ref.~\cite{fort}, the laser beam
is considered with a detuning of 67 nm to the red of the 5 $^2S_{1/2}$
- 5 $^2P_{1/2}$ transition at 795 nm and an irradiance of 6 mW/$\mu$m$^2$
that determines the value of the coupling constant
$\vert \vec\mu_{12}\cdot E_0\hat e_x\vert/\hbar$. The trajectories of
the atoms are described taking  the laser wavelength as unit of
length and, as unit of time, the inverse of the Einstein coefficient
$\Gamma$ which, for 5$^2P_{1/2}$ state of $^{85}$Rb, is 3.7 $\times
10^{7}s^{-1}$. The initial kinetic energies are reported in terms of
the corresponding ``temperature" by dividing by the Boltzmann constant $k_B$.
Although we have analyzed several values of the light field characteristic
parameters, in order to be specific we report just the results where
$k_z=0.995\omega/c$ and the topological charge $m=2$. This makes a paraxial
realization of the beam a good approximation, and admits the possibility of
observing light-atom angular momentum transfer.\\
In  the reported clouds, the range of initial conditions of the atoms is: $0.01\lambda\leq\rho\leq 2.6\lambda$; $0.0001\lambda \leq
z \leq 0.001\lambda$; $-0.0001\lambda\Gamma\leq
\dot{\rho}\leq 0.0001\lambda\Gamma$;
$0.0001\Gamma\leq\dot{\varphi}\leq
0.00015\Gamma$; $-0.0025\lambda\Gamma\leq \dot{z} \leq
-0.001\lambda\Gamma $ with the the initial kinetic energies ranging
from $\sim 5\mu K$ to $\sim 30\mu K$.

\vspace{0.5cm}
{\bf Case 1}. Single rotating Bessel beam.

The optical potential energy linked to the conservative factor $\vec
\alpha$, in this case consists of annular potential wells,
corresponding to the concentric bright rings of the intensity
distribution of the BB. An atom trapped in one of the bright rings
of the beam oscillates in the radial direction around the minimum of
the potential energy with an amplitude that depends on its initial
position and velocity.
 Along the $z$ axis, the
atom is subjected to the gravity force and to the dissipative term
$\vec \beta$ of the optical force, associated with the phase of the
light field $\vec\beta=\vec\nabla\phi$. In a FORT with the
parameters given above, in this case, the dissipative effects have small influence
on the atom's motion. Hence, the gravity force dominates the evolution
along the z-axis and the transfer of orbital angular momentum from the
beam to the atom is also negligible. If the detuning were smaller, the effect of the
dissipative forces would be larger, and the optical acceleration
along the azimuthal and axial directions would eventually become
noticeable.

\begin{figure}
\includegraphics[width=3in]{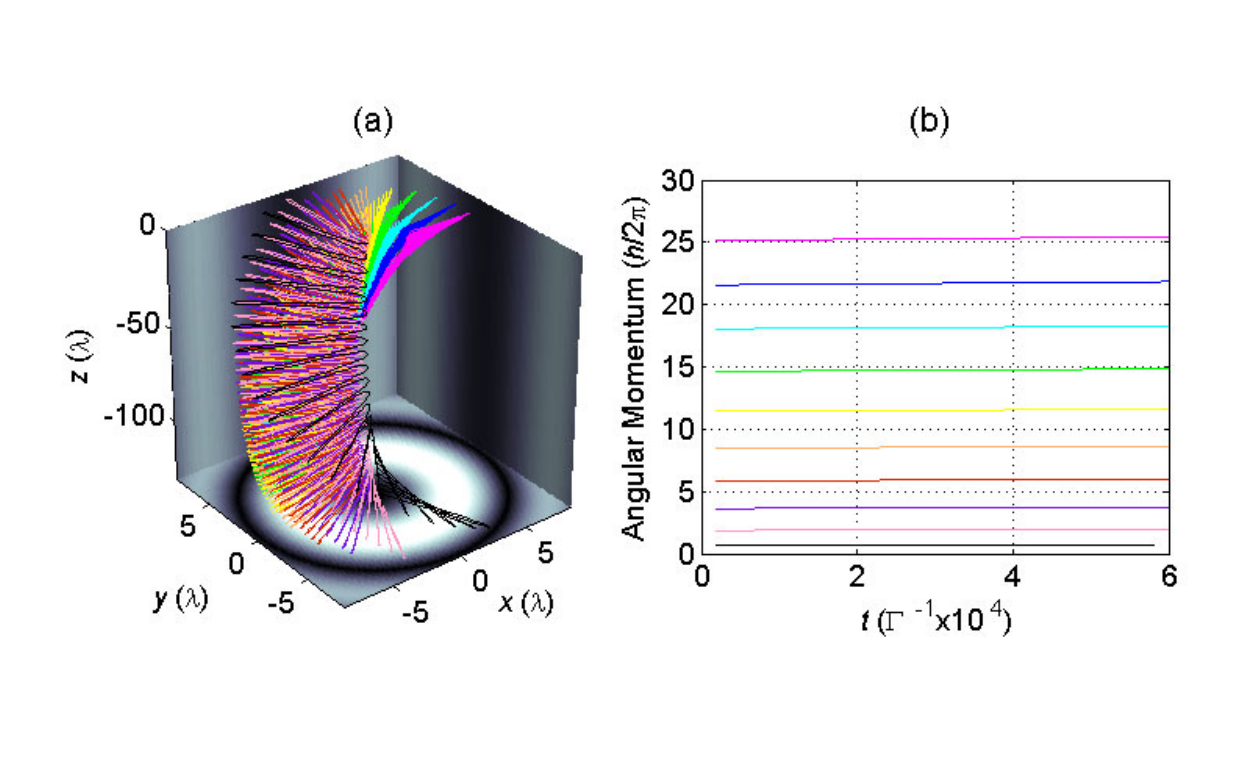}
\caption{(a) Spatial paths of ten atoms representing the atomic
cloud described in the text moving within a rotating BB. The
topological charge of the BB is $m=2$ (positive helicity), and it is
propagating along the positive $z$ axis (upwards direction); the
axial and transverse components of its wave vector are $k_z=0.995k$
and $k_{\bot}=0.0999k$. The wavelength of the light is
$\lambda=862$nm, which is the length unit in the plots. The smallest
value of the kinetic energy and the largest absolute value of the
potential energy correspond to the magenta path, whereas the
opposite occurs for the black path. The starting point of all the
paths is the plane $z\simeq0$. Notice the scale differences between
the three spatial axes. (b) Angular momentum as a function of time
for the same atoms. The range of initial conditions of the atoms in
the analyzed cloud is:  $0.01\lambda\leq\rho\leq 2.6\lambda$;
$0.0001\lambda \leq z \leq 0.001\lambda$;
$-0.0001\lambda\Gamma\leq \dot{\rho}\leq
0.0001\lambda\Gamma$; $0.0001\Gamma\leq\dot{\varphi}\leq
0.00015\Gamma$,$-0.0025\lambda\Gamma\leq \dot{z} \leq
-0.001\lambda\Gamma )$ with the the initial kinetic energies ranging
from $\sim 5\mu K$ to $\sim 30\mu K$. The colors of the paths in (a) are
directly correlated to the angular momentum in (b).}\label{caso1-ma}
\end{figure}

Figure~\ref{caso1-ma} illustrates the typical trajectories for ten
atoms taken from the atomic cloud described above and downloaded in
a second order BB at $z\simeq 0$.   Although the light beam is propagating
upwards, the atoms move downwards due to their initial negative
velocities and to the acceleration of gravity. Gravity also helps to
keep the atoms stably trapped within the beam profile in the
transverse direction, even if the initial radial position of an atom
is close to the axial node ($\rho \sim 0.01 \lambda$). In contrast, we have
verified that some atoms would escape in the radial direction if the
atomic cloud were considered with similar initial conditions
 but with positive sign of the z-component of the velocity.
 In Fig.~\ref{caso1-ma}(b), we
also show the angular momentum as a function of time for each of the
atoms illustrated in Fig.~\ref{caso1-ma}(a). In all the cases, the
angular momentum remains practically equal to its initial value.\\

Single rotating BBs have been proposed before as guides for cold
atoms \cite{ArltHD00, ad1} though no semiclassical calculations
were reported. Here, we have considered a red detuned system and
found that a BB may indeed be used as an atom guide, in this case. \\

\vspace{0.5cm}
{\bf Case 2}. Twisted helical lattice.

The intensity distribution for this case is illustrated in
Fig.~\ref{fig2}. A transverse cross section looks like a stationary
BB, but it is rotating as a whole along the $z$ axis, completing a
revolution in a distance of $\left|m\right|\lambda_z$. For a red
detuned lattice, the potential energy minima correspond to a set of
$2\left|m\right|$ twisted intertwined pipes between each pair of
radial nodes, along which the atoms can be guided in independent
channels. This light configuration is analogous to that proposed by
Bhattacharya \cite{twist} for LG beams. However, for BB beams, the
propagation invariance introduces additional features for atom
guiding with respect to LG beams.

\begin{figure}
\includegraphics[width=3in]{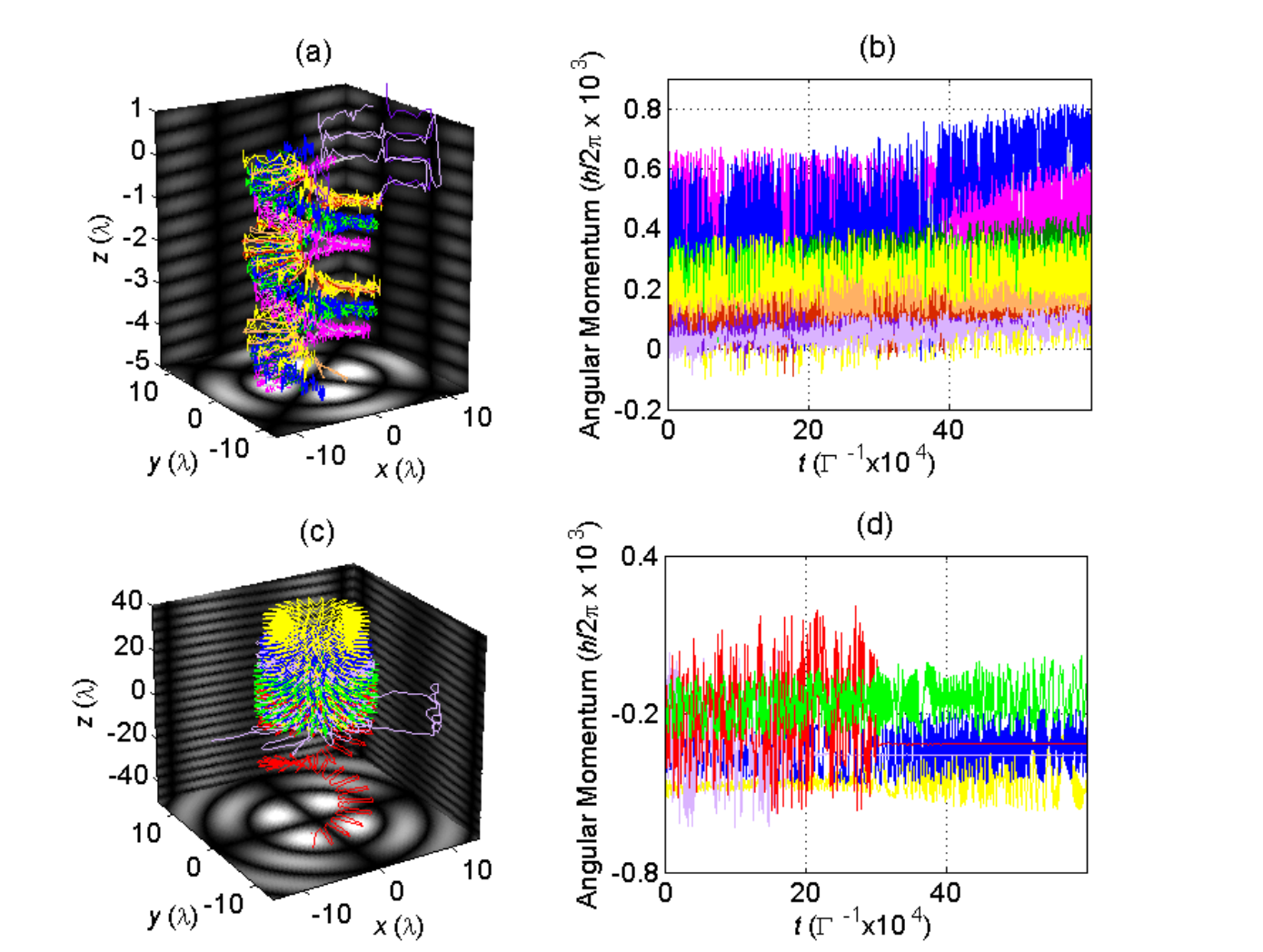}
\caption{(a) Illustrative examples of the spatial trajectory  and
(b) angular momentum of atoms moving within a twisted helical Bessel
lattice generated in the way described in Fig.~2. The parameters of
the two superimposing beams and the range of initial conditions of
the atoms in the analyzed cloud are the same as those of
Fig.~\ref{caso1-ma} so that the atoms motion is initially  downwards
and has the same helicity as the light pattern. In Figure (a) the
time interval corresponds to $0<T<0.5
\times 10^4\Gamma^{-1}$. Figure (c) shows illustrative examples of the spatial trajectory
and (d) angular momentum of atoms moving within the same twisted Bessel
lattice than in (a); the initial conditions of
the atoms in the analyzed cloud are the same as those used in
(a) but with the angular velocity $\dot\varphi$
reversed. Notice the scale difference in the axial coordinate in figures (a) and (c).
The colors of the paths in (a) and (c) are directly
correlated to the angular momentum in (b) and (d).
}\label{fig_cuerda}
\end{figure}

Since the axial and azimuthal variables appear in the combination
$(m\varphi+k_zz)$, the light field amplitude has a well defined
helicity. In absence of gravity, it is expected that an atom
initially moving with the same helicity than that of the light
pattern will preserve it, although its angular momentum may change
in magnitude. Otherwise, i. e., if an atom has an initial motion with
different helicity than that of the light pattern, the optical force
might be able to change the atoms helicity. We have verified this
fact numerically taking into account gravity effects. This is illustrated in
Fig.~\ref{fig_cuerda} where: (a) corresponds to loading an
atom cloud at the $z\simeq 0$ plane with the parameters mentioned
above, so that the atoms initially move downwards and
the atoms and the light pattern have the same
helicity; (c) corresponds to loading an
atom cloud also downwards  with the same parameters
mentioned above with the exception of $\dot \varphi$ whose sign has been reversed, $-0.0015\Gamma\leq\dot{\varphi}\leq -0.001\Gamma$, so that
initially, atoms  and  light pattern have
opposite helicity. In the latter case, the light force attempts to change the
helicity of the atoms by sending them upwards acting
against gravity; when it is not able to do so, the atom exhibits
a complicated trajectory that may end with its escape. Notice that
in both  cases (a) and (c), some atoms are able to escape from the
radial confinement in the Bessel ring,
in particular when they are initially  located close to or
at a nodal surface; however, they may be eventually trapped at higher radii.
The time dependent angular momentum of each atom in both clouds has strong
oscillations with an increasing average value, as illustrated in Fig.~\ref{fig_cuerda}
(b) and (d).
  For atoms with a helicity coinciding with that of the light pattern, an angular
momentum average that starts being $~10\hbar$, as in Case 1, ends
with values up to $~5\times10^2\hbar$ for $t\sim 10^5 \Gamma^{-1}$.

 These results show that: (i) twisted helical beams act as wave guides
 with intertwined channels that determine the rotation direction of radially
 trapped atoms, and (ii) a significant amount of angular momentum can be
transferred to atoms using this beam configuration.\\

\vspace{0.5cm}
{\bf Case 3}. 3D stationary circular lattice.

\begin{figure}
\includegraphics[width=3in]{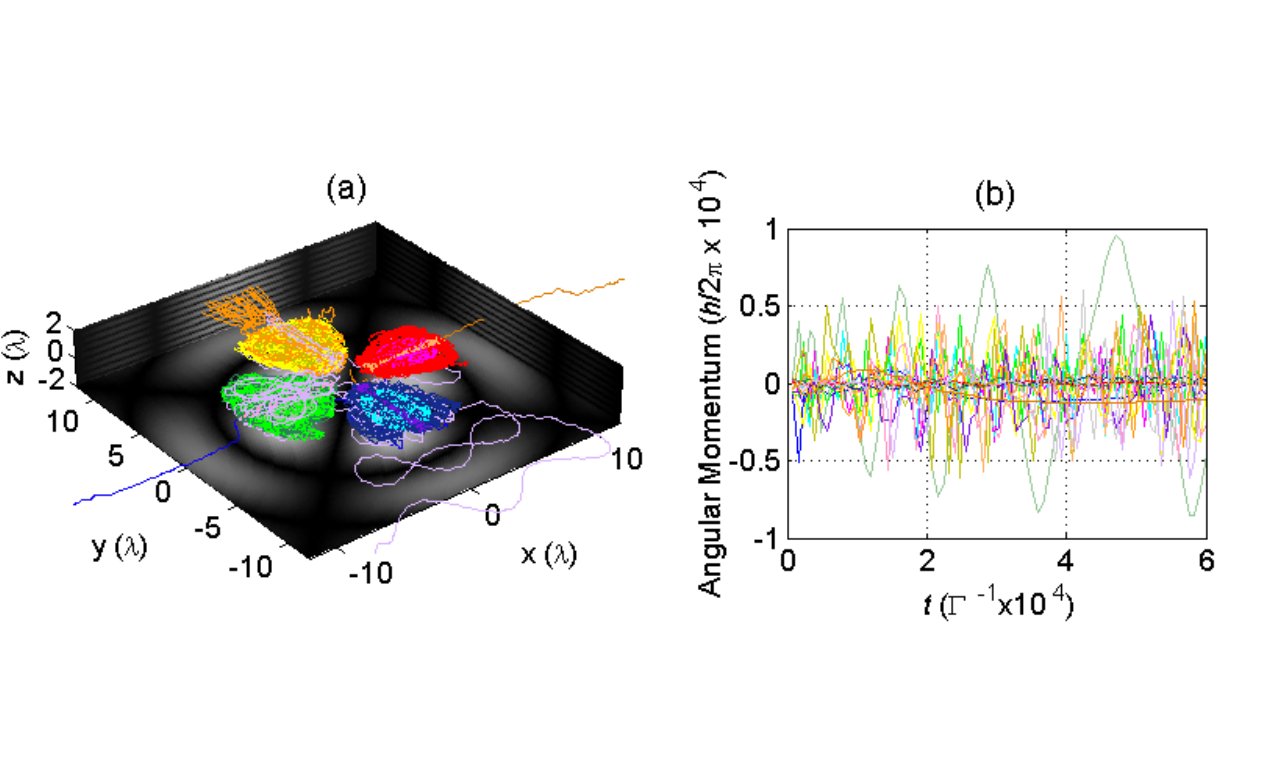}
\caption{(a) Spatial path of atoms moving within a 3D Bessel lattice
generated by the superposition of two stationary BBs of second order
propagating in opposite directions, as described in Fig.~3. (b)
Angular momentum as a function of time for the same atoms. The
parameters of the two superimposing beams and the range of initial
conditions of the atoms in the analyzed cloud are the same as those
of Fig.~\ref{caso1-ma}.  The colors of the paths in (a) are directly
correlated to the angular momentum in (b).}\label{Fig_est3D}
\end{figure}

The intensity distribution of this lattice (Fig.~3) corresponds to a
set of individual \textquoteleft potential cages' (potential wells in all the
three spatial dimensions) distributed around in a coordinate system
with circular cylindrical geometry. Nodal surfaces define the limits
of the potential cages. The numerical simulation with the cloud
described above shows that: (i) atoms initially located at a nodal
surface have a high probability of escaping from the lattice due to
a lack of potential energy, particularly when they are very close to the
 axis of symmetry of the beam; (ii) for atoms initially within a
cage, so that their total initial energy is negative, the trapping in
this lattice results very robust, regardless of the direction of its
initial momentum and in spite of the presence of gravity. In some
cases, depending on its initial position and velocity, an atom may
tunnel from one cage to the next one either along the axial or the
azimuthal direction, while it keeps confined in the other
directions. In the latter case, the atom may stay trapped in a
transverse plane, going around the whole beam circumference.
Illustrative examples of trajectories for the atomic cloud described
above are given in Fig.~\ref{Fig_est3D}.

In Fig.~\ref{Fig_est3D}, we also observe large fluctuations of
the orbital angular momentum $L_z$.  By comparing with the results for the twisted helical
configuration, Fig.~\ref{fig_cuerda}, we notice that the stationary
condition in angular and axial directions increases the efficiency
of angular momentum transfer in at least one order of magnitude.
 Notice that in this case, the mean  axis of rotation of each
atomic trajectory is located
in the cages, so that it does not coincide with the axis of the beam.

Based on the high angular momentum oscillations, we consider that this
kind of lattice might be especially useful in the generation of
vortices in degenerate gases. Notice as well that it would be a more
appropriate choice for the studies of quasi one dimensional systems
with periodic boundary conditions along the azimuthal direction
proposed by Amico and coworkers \cite{amico}, since the scheme they
proposed of interfering a plane wave with a Laguerre-Gaussian beam
would give rise to spiral fringes \cite{revolving} rather than
localized spots as in this case. Furthermore, in a 3D stationary
Bessel lattice the axial confinement is achieved all-optically
instead of magneto-optically. In addition, this lattice could  be
also a suitable choice for studies of atomic wave function
interference between components
that rotate in opposite directions.\\

\vspace{0.5cm}
{\bf Case 4}. Toroidal train lattice.

\begin{figure}
\includegraphics[width=3in]{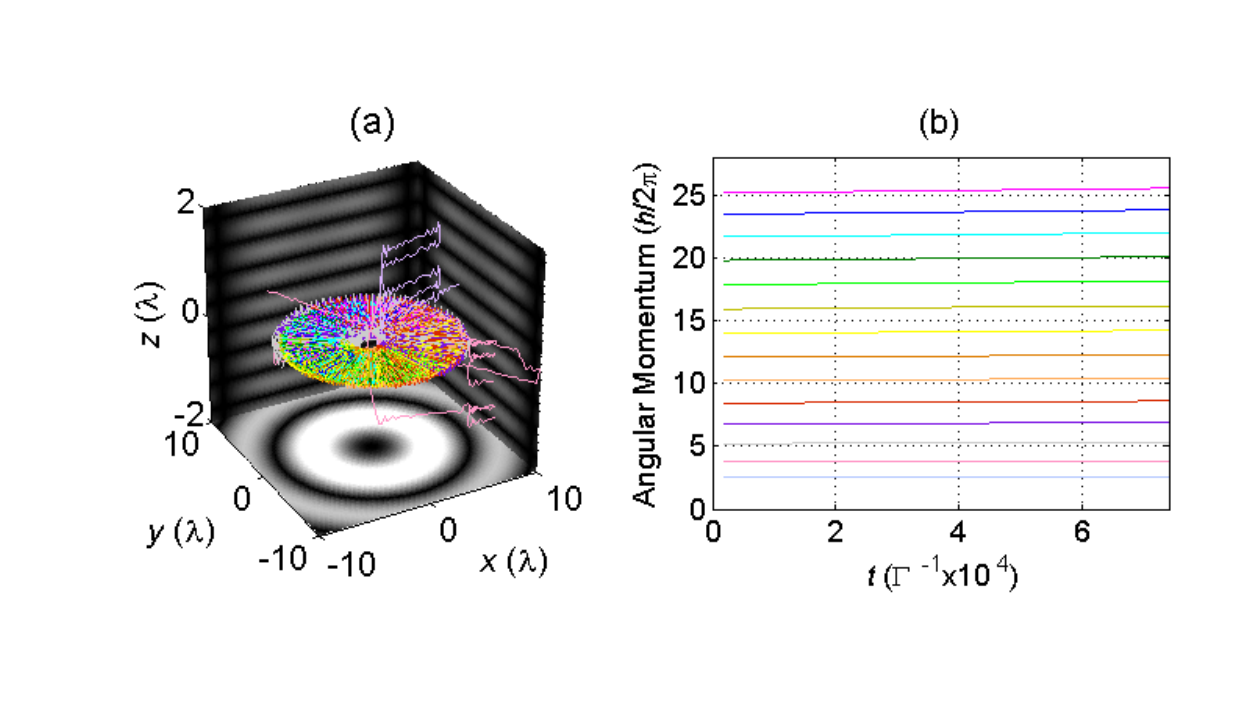}
\caption{(a) Spatial path of atoms moving within a toroidal Bessel
lattice generated by the superposition of rotating BBs of second
order propagating in opposite directions, as described in
Fig.~\ref{fig_4}.(b) Angular momentum as a function of time for the
same atoms. The parameters of the two superimposing beams and the
range of initial conditions of the atoms in the analyzed cloud are
the same as those of Fig.~\ref{caso1-ma}. The colors of the paths in
(a) are directly correlated to the angular momentum in
(b).}\label{fig_donas}
\end{figure}

The optical potential energy in this case corresponds to a set of
toroidal cages aligned along the $z$ axis; the intensity pattern is
shown in Fig.~\ref{fig_4}.  In
general, an atom initially located at an antinodal $z$ plane
($z=n\lambda_z/2$)
will remain trapped in a single torus, waving
along the radial and axial directions, and rotating around the beam
axis, provided its total initial energy is negative. On the other
hand, atoms initially located at a nodal surface may hop to neighbor
toroidal traps either one way or the other, while keeping trapped in
the radial direction and rotating around the beam axis. This hopping
behavior may be exhibited during relatively long time intervals
($t\approx10^5\Gamma^{-1}$) before the atom finally escapes. Typical
examples of paths followed by atoms of a cloud trapped in this
lattice can be seen in Fig.~\ref{fig_donas}(a).
If an atom has a non
null azimuthal component of its initial velocity, it will remain
rotating around the beam axis at practically constant average
angular velocity. This can be appreciated from
Fig.~\ref{fig_donas}(b), where the angular momenta of the different
atoms remain almost constant and have the same order of magnitude
than in the case 1, of the propagating rotating BB, which is much
smaller than in the other two cases studied here.

\section{3D atom circuits with combination of Bessel lattices}
With a clear picture of the mechanical behavior of  atoms in the
different light fields we have discussed so far, we are in a
position to elucidate a more sophisticated application. By
alternating the operation of different lattices in an appropriated
combination, it is possible to create what we will call
\textquoteleft atom loops\textquoteright. These loops can either be
open or \textquoteleft closed \textquoteright. By
a \textquoteleft closed'  loop we do not mean, of course, that the atom will
come back to its initial position, but rather to approximately the
same spatial region.

For instance, consider a cloud of atoms downloaded in a toroidal train lattice.
After a transitory time, we obtain a steady cloud of atoms trapped in the radial and axial
direction moving with essentially their initial angular momentum. By applying the
twisted helical BB,
they will move downwards or upwards depending on the helicity of the beam and the direction
of the atomic azimuthal velocity. In general,  the twisted helical BB preserves the radial trapping
 and modifies the atomic angular momentum $L_z$. Now, by
turning on a toroidal train lattice just after the twisted helical BB is turned off, the atoms
will again be axially trapped. If most of the atoms in the first
toroidal lattice moved nearby the $z=0$ plane, in the final configuration, we expect that
most atoms rotating in toroidal cages with $z>0$  will have an opposite
angular momentum $L_z$ to those rotating in cages with $z<0$.  We confirmed
these ideas by performing several numerical simulations of the process. For instance, consider an atomic cloud with an initial average kinetic energy $<K_{in}>\approx10\mu$K and an initial angular momentum average $<\vert L_z\vert> \approx50 \hbar $. The twisted helical BB is applied during a time interval $\Delta T =5 \times 10^4 \Gamma^{-1}$. In the final configuration, in each toroidal cage,  $\sim 85\%$ of the  trapped atoms had a common direction of rotation about the $z$ axis. This direction was opposite for $z>0$ and for $z<0$. During the process, $10\%$ of the atoms were radially lost.

With the current technology of spatial light modulators (SLM), the switching among different options of optical lattices may be performed at reasonably high speeds, limited only by the response time of the specific light modulation device. An experimental study on the interactive generation and switching of the light patterns studied here will be reported elsewhere. Here, we assume valid a sudden approximation in which the atoms do not modify their state of motion during the switching.

We analyzed other loops. As expected, in all cases, the higher the number
 of steps to obtain a predesigned
path, the lower the number of atoms in a cloud that are able to
follow it. This can become an advantage of the procedure when the
purpose is to select atoms with predetermined mechanical parameters.

\begin{figure}
\includegraphics[width=3in]{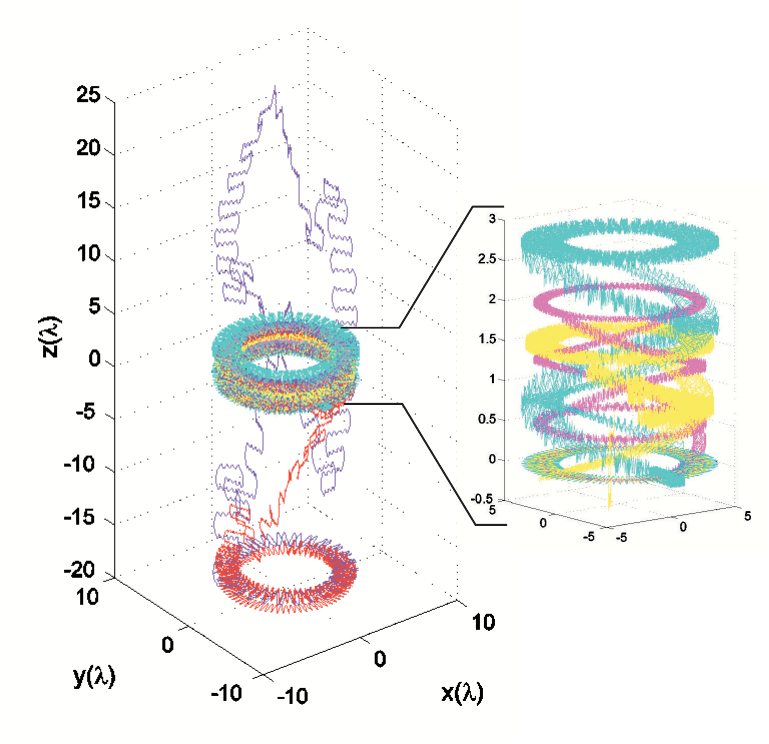}
\caption{ Some illustrative closed and open atom loops obtained by the effect
of five different light fields operated consecutively: (1) toroidal; (2) twisted helical;
(3) toroidal; (4) twisted helical with opposite helicity than in (2); (5) toroidal.
  The general parameters of the light fields correspond to those in Fig.~(\ref{fig_cuerda}) and (\ref{fig_donas}).  The initial average kinetic energy  is $<K_{in}>=10\mu$K
 and the initial angular momentum is $ <\vert L_z\vert > ~50 \hbar $.
 The application times of each step were optimized to obtain a closed loop for
atoms with initial conditions close to those of the blue trajectory. Most of the atoms
performed trajectories like the red or purple ones, that is, open loops since their
final trajectory is not close to the initial one. However, radial looses were not too
frequent during the whole procedure (about $15\%$) and most of them occur in the first three steps.}\label{figcircuito}
\end{figure}

Let us consider the following five step circuit:
step~(1) an atom cloud is trapped in a
toroidal cage for a given time interval; step~(2) a twisted helical BB is applied with
the proper helicity to send the atoms upwards (downwards)
if $\dot\varphi>0$ ($\dot\varphi<0$); step~(3) their are trapped again
in a toroidal cage; step~(4) they are sent downwards (upwards) using a twisted helical
BB with opposite $k_z$ than in step (2); step (5) their are trapped again by a toroidal cage.
This circuit admits the possibility of obtaining closed atomic loops. This is illustrated in
Fig.~\ref{figcircuito} for an atomic cloud that had an initial average kinetic energy  $<K_{in}>\approx10\mu$K and an initial angular momentum $ <\vert L_z\vert > \approx 50 \hbar $. The application
 time intervals for each step were optimized to obtain a closed loop for a small, $~5\%$, subset of atoms  corresponding to those with the larger initial radial position ($R\sim6\lambda$). In the procedure, we observed that most radial loses occur at the first three steps. After that about $~10\%$  of the trajectories corresponded to closed loops. All those atoms had the same direction of rotation.

Figure~\ref{figcircuito} illustrates just an example of a loop,
but other combinations can be explored for different purposes.
Notice that atoms with preselected energies and  momenta
  could be guided in different directions, so that circuits
  could be designed with the possibility of performing  atom interference experiments.

\bigskip

\section{Conclusions}
We have analyzed the dynamical behavior of an atomic cloud moving under the
action of four different configurations of light fields with
circular cylindrical geometry: a propagating-rotating Bessel beam of
order $m$, a twisted helical
lattice or twisted helical field, a 3D stationary circular lattice,  and a toroidal train lattice.
We presented the fundamental equation for the optical force,
 based on references \cite{gordon-ash} and \cite{leto}, and gave the specific
expressions for the conservative ${\vec \alpha}$ and dissipative ${\vec\beta}$
terms of the force in each of the cases under study. In section 4, we discussed the
numerical results, case by  case, of the solution of the motion equations for the atom, for a red detuned  far-off-resonance system. We found that the single rotating Bessel beam and
 the twisted helical lattice can be used to guide atoms, in the latter case along
 $2\vert m \vert$ separate channels, whereas the 3D stationary circular lattice and
 the toroidal train lattice can be applied to obtain 3D confinement within a small
 region of the space. The twisted helical lattice can be used to select atomic helicities
 and gives rise to strong angular momentum oscillations. The 3D stationary circular
  lattice define a mean rotation axis for the atomic trajectories located at
   each potential cage. Finally, on the basis of our numerical
 results, we proposed an application consisting of the consecutive operation of the
 different options of light fields studied here, in order to create atom loops in predesigned ways by all-optical means.\\

Even when we have restricted our analysis here to the case of Bessel
modes, it is worth to appreciate that, in the paraxial versions, all
the lattices or light fields discussed above would have an analogous
in terms of Laguerre-Gaussian  laser modes, which might be easier to
generate experimentally. In that case, however, beam spreading on
propagation should be taken into account; the waist plane of the
interfering beams should coincide and the alignment may become an
issue. In general, regardless of the specific form of the radial
profile, any beam with circular cylindrical symmetry could be useful
for generating similar lattices to those studied here, and the behavior of cold
atoms in such lattices is expected to be also analogous to that
discussed in this paper. It is worth to mention as well, that in the
specific case of Bessel lattices with light irradiance of about 6
mW/$\mu$m$^2$, we found that atoms with initial kinetic energies up
to $30 \mu$K can be trapped not only in the first ring of the Bessel
profile, but also in the second and even in the third outer rings.

There are also very interesting features
occurring in the cases of near-resonance conditions and blue-detuned
systems, that are by themselves worthy of other thorough studies.
For instance, Gommers and coworkers considered near-resonance conventional
lattices to experimentally generate an atomic ratchet \cite{ratchet}; the lattices
 studied here may represent very attractive novel options for this purpose,
 since one can generate quasi periodic systems with open boundary conditions along the $z$
axis, but also with closed open boundary conditions along the azimuthal
coordinate, as well as explore ratchet systems with new geometries.
 Thus, a possibility for performing novel studies on
 stochastic dynamics in atomic systems is opened both classically and quantum mechanically
 since atomic temperatures will define the proper dynamics. A quantum description of the
dynamics of atoms in cylindrical beams will be reported elsewhere.\\

{\bf Acknowledgements.} R. J. thanks useful discussions to Prof. J.
Rec\'amier. This work was partially supported by DGAPA-UNAM,
projects IN103103 and IN115307.

\end{document}